\documentclass[12pt,a4paper]{article}
\usepackage{graphicx}%
\usepackage{amsmath}
\usepackage{amsfonts}%
\usepackage{amssymb}
\usepackage{epsfig}

\begin{document}

\begin{center}
{\bfseries Scalar $\sigma$ meson at finite temperature in nonlocal quark model
\footnote{{\small Talk at the Round Table Discussion "Searching for the mixed phase of
strongly interacting matter at the JINR Nuclotron", JINR, Dubna, July 7 - 9, 2005.}}}

\vskip 5mm

D. Blaschke$^{\dag,\ddag}$, Yu. L. Kalinovsky$^\S$, A. E. Radzhabov$^\dag$ and M. K.
Volkov$^\dag$

\vskip 5mm

{\small {\it $^\dag$ Bogoliubov Laboratory of Theoretical Physics, \\
Joint Institute for Nuclear Research, Russia}} \\
{\small {\it $^\ddag$ Theory Division, GSI mbH, D--64291 Darmstadt, Germany}}
\\
{\small {\it $^\S$ Laboratory of Information Technologies,\\ Joint Institute for
Nuclear Research, 141980 Dubna, Russia}}
\\

\end{center}

\vskip 5mm

\begin{center}
\begin{minipage}{150mm}
\centerline{\bf Abstract} Properties and temperature behavior of $\pi$ and $\sigma$ bound
states are studied in the framework of the nonlocal model with a separable interaction kernel
based on the quark Dyson-Schwinger and the meson Bethe-Salpeter equations. $M_\pi(T)$,
$f_\pi(T)$, $M_\sigma(T)$ and $\Gamma_{\sigma \to \pi\pi}(T)$ are considered above and below
the deconfinement and chiral restoration transitions.
\end{minipage}
\end{center}

\vskip 10mm

\section{\label{sec:intro}Introduction}

Understanding the  behavior of matter under extreme conditions is nowadays a challenge in the
physics of strong interactions. Different regions of the QCD phase diagram are an object of
interest, and major theoretical and experimental efforts have been dedicated to the physics
of relativistic heavy--ion collisions looking for signatures of the quark gluon plasma [QGP]
\cite{a1,a2,a3}.

Restoration of symmetries and deconfinement are expected to occur at high-density and/or
temperature. In this regard, the study of observables of pseudoscalar and scalar mesons is
particularly important. Since the origin of these mesons is associated with the phenomena of
spontaneous and explicit chiral symmetry breaking, its temperature behavior is expected  to
carry relevant signs of a possible restoration of symmetries. Usually, the restoration of
chiral symmetry at high temperature is connected with the transition of hadron matter into
quark-gluon plasma.

Effective quark models are useful tools to explore the behavior of matter at temperatures.
Nambu--Jona-Lasinio [NJL] \cite{njl} type models have been extensively used over the past
years to describe low-energy features of hadrons and also to investigate restoration of
chiral symmetry with temperature \cite{njlT1}-\cite{njlT4}.

This paper is devoted to investigation of the phase transition in hot matter and the
temperature behavior of pseudoscalar as well as scalar mesons in the framework of the
effective nonlocal model. This work is the continuation of \cite{ijmpa}. In \cite{ijmpa}, a
special separable form of effective gluon propagator is used in construction of the quark
Dyson - Schwinger equation (DSE) and the Bethe - Salpeter equation (BSE) for bound states.
Only pseudoscalar and vector mesons are considered in that paper. Here we concentrate on the
properties of the scalar $\sigma$ meson at finite temperature.

\section{Dyson - Schwinger equation with separable interaction}
The dressed quark propagator $S(p)$ and meson Bethe - Salpeter (BS)
amplitude $\Gamma(p,P)$ are solutions of the DSE
\cite{cr1}-\cite{cr4}
\begin{eqnarray}\label{sde}
  S(p)^{-1} = i \hat{p} + m_0 +
  \frac{4}{3} \int \frac{d^4q}{(2\pi)^4} g^2 D_{\mu\nu}^{\mbox{eff}} (p-q)
  \gamma_\mu S(q) \gamma_\nu \label{DSE}
\end{eqnarray}
and the BSE equation
\begin{eqnarray}\label{bse}
  \Gamma(p,P) = \frac{4}{3} \int \frac{d^4q}{(2\pi)^4} g^2 D_{\mu\nu}^{\mbox{eff}} (p-q)
  \gamma_\mu S(q_+) \Gamma(q,P)  S(q_-) \gamma_\nu \label{BSE},
\end{eqnarray}
where\footnote{We use the Euclidean metric.} $D_{\mu\nu}^{\mbox{eff}} (p-q)$ is an "effective
gluon propagator", $m_0$ is the current quark mass, $P$ is the total momentum, and
$q_{\pm}=q\pm P/2$. The form of equations (\ref{DSE}) and (\ref{BSE}) corresponds to  the
rainbow - ladder truncations of DSE and BSE.

The simplest separable Ansatz $g^2 D_{\mu\nu}^{\mbox{eff}} (p-q) \rightarrow
\delta_{\mu\nu} D(p^2,q^2,p\cdot q)$ in a Feynman - like gauge is employed
\begin{eqnarray}%\label{}
D(p^2,q^2,p\cdot q) = D_0 f_0(p^2) f_0(q^2) + D_1 f_1(p^2) (p\cdot q) f_1(q^2)
\end{eqnarray}
This is a rank-2 interaction with two strength parameters $D_0$, $D_1$ and the corresponding
form factors $f_i(p^2)$. The choice for these quantities is constrained to the solution of
the DSE for the quark propagator (\ref{DSE})
\begin{eqnarray}%\label{}
  S(p)^{-1} = i \hat{p} A(p^2) + B(p^2)
\end{eqnarray}
with $B(p^2) = m_0 + b f_0(p^2)$ and $A(p^2)=1 + a f_1(p^2)$, where $a$ and $b$ are some
constants. If there are no poles in the quark propagator $S(p)$ for real timelike $p^2$, then
there is no physical quark mass shell and a meson cannot decay into quarks. The propagator is
confining \footnote{A similar confining propagator without poles is used in a nonlocal model
of NJL type\cite{RV1,RV2}.} if $m^2(p^2) \neq -p^2$ for real $p^2$ where the quark mass
function is $m(p^2)=B(p^2)/A(p^2)$. It is found that the simple choice \mbox{$f_i(p^2) =$}
\mbox{exp $(-p^2/\Lambda_i^2)$} leads to a reasonable description of $\pi$ and $\sigma$ meson
properties. At the same time, the produced quark propagator is found to be confining and the
infrared strength and shape of the quark amplitudes $A(p^2)$ and $B(p^2)$ are in qualitative
agreement with the results
of typical DSE studies. %~\cite{MR97}.
We use the exponential form factors as a minimal way to preserve these properties while
realizing that the ultraviolet suppression is much greater than the power law fall-off
(with logarithmic corrections) known from asymptotic QCD. The total number of model
parameters is five.

The extension of the separable model studies  to the finite temperature case, $T\neq 0$, is
systematically accomplished by a transcription of the Euclidean quark 4 - momentum via {$q
\rightarrow$} {$ q_n =$} {$(\omega_n, \vec{q})$}, where {$\omega_n=(2n+1)\pi T$} are the
discrete Matsubara frequencies. The effective $\bar q q$ interaction will automatically
decrease with increasing $T$ without the introduction of an explicit $T$-dependence which
would require new parameters.

The result of the DSE solution for the dressed quark propagator now becomes
\begin{eqnarray}
S^{-1}(p_n, T) = i\vec{\gamma} \cdot \vec{p}\; A(p_n^2,T)
                            + i \gamma_4 \omega_n\; C(p_n^2,T)
                            + B(p_n^2,T) \; ,
\label{invprop}
\end{eqnarray}
where {$p_n^2=\omega_n^2 + \vec{p}^{\,2}$}. The solutions have the form {$B= m_0 + b(T)
f_0(p_n^2)$}, \mbox{$A=1+ a(T) f_1(p_n^2)$}, and {$C=1+ c(T) f_1(p_n^2)$}, and the DSE
becomes a set of three non-linear equations for $ b(T)$, $a(T)$, and $c(T)$.  The explicit
form is
\begin{eqnarray}
 a(T) &=& \frac{8 D_1}{9}\,  T \sum_n \int \frac{d^3p}{(2\pi)^3}\,
 f_1(p_n^2)\, \vec{p}^{\,2}\, [1 +  a(T) f_1(p_n^2)]\; d^{-1}(p_n^2,T) \; ,
\\
 c(T) &=& \frac{8 D_1}{3}\,  T \sum_n \int \frac{d^3p}{(2\pi)^3}\,
 f_1(p_n^2)\, \omega_n^2\, [1 +  c(T) f_1(p_n^2)]\;
                                                d^{-1}(p_n^2,T) \; ,
\\
 b(T) &=& \frac{16 D_0}{3}\,  T \sum_n \int \frac{d^3p}{(2\pi)^3}\,
 f_0(p_n^2)\, [m_0 +  b(T) f_0(p_n^2)]\; d^{-1}(p_n^2,T) \; ,
\end{eqnarray}
where $d(p_n^2,T)$ is given by
\begin{eqnarray}
d(p_n^2,T) = \vec{p}^{\,2}\, A^2(p_n^2,T) +\omega_n^2\,  C^2(p_n^2,T)
                  + B^2(p_n^2,T) .
\end{eqnarray}

\section{Pseudoscalar and scalar correlations}

With the separable interaction  the
allowed form of the solution of (\ref{bse})  for the pion BS amplitude
$\vec{\tau}\, \Gamma_\pi(q;P)$ is
\begin{eqnarray}
\Gamma_\pi(q;P) =\gamma_5 \left(i E_\pi (P^2) +
                     \hat{P} F_\pi(P^2)\right) \; f_0(q^2) \; .
\label{pibsa}
\end{eqnarray}
The $q$ dependence is described only by the first form factor $f_0(q^2)$.  The second term
$f_1$ of the interaction can contribute to BS amplitude only indirectly via the quark
propagators. The pion BSE, (\ref{BSE}),  becomes a $2\times 2$ matrix eigenvalue problem
\mbox{${\cal K}(P^2) f = \lambda(P^2) f$} where the eigenvector is \mbox{$f = (E_\pi,
F_\pi)$}.   The kernel is
\begin{eqnarray}
{\cal K}_{ij}(P^2) = - \frac{4 D_0}{3}\, {\rm tr_s}\, \int\,
       \frac{d^4q}{(2\pi)^4}f_0^2(q^2)
 \left[ \hat{t}_i\, S(q_+)\, t_j\, S(q_-)\,  \right]~,
\label{pikernel}
\end{eqnarray}
where the $\pi$ covariants are \mbox{$t=(i\gamma_5, \gamma_5\, \hat{P})$} with
\mbox{$\hat{t}=(i\gamma_5,$} \mbox{$-\gamma_5\, \hat{P}/2P^2)$}. We note that the separable
model produces the same $q^2$ shape for both amplitudes $F_\pi$ and $E_\pi$; the shape is
that of the quark amplitude $B(q^2)$. Goldstone's theorem is preserved by the present
separable model; in the chiral limit, whenever a nontrivial DSE solution for $B(p^2)$
exists, there will be a massless $\pi$ solution to (\ref{pikernel}).

The normalization condition for the pion BS amplitude can be expressed as
\begin{eqnarray}
\label{pinorm}
    \left. 2 P_\mu = 2 N_c \,  \frac{\partial}{\partial P_\mu} \,
        \,  \int \frac{d^4q}{(2\pi)^4}\, {\rm tr}_s \left(
        \bar\Gamma_\pi(q;-K)\, S(q_+)\,
        \Gamma_\pi(q;K)\,S(q_-) \right)
        \right|_{P^2=K^2=-M_\pi^2} \,.
\end{eqnarray}

Here $\bar\Gamma(q;K)$ is the charge conjugate amplitude $[{\cal C}^{-1} \Gamma(-q,K) {\cal
C}]^{\rm t}$,  where ${\cal C}=\gamma_2 \gamma_4$ and the index t denotes a matrix transpose.
The pion decay constant $f_\pi$ can be expressed as the loop integral
\begin{eqnarray}
   f_\pi \; P_\mu\,\delta_{ij} &=& \langle 0|\bar q \frac{\tau_i}{2}
   \gamma_\mu \gamma_5 q | \pi_j(P)\rangle
\nonumber \\
        &=& \delta_{ij} \, N_c \, {\rm tr_s} \int \frac{d^4q}{(2\pi)^4}\;
     \gamma_5 \gamma_\mu \; S(q_+) \; \Gamma_\pi(q;P)\; S(q_-)\; .
\label{fpi}
\end{eqnarray}

For a chiral partner of the pion, $\sigma$ meson, we take the BS amplitude with only one
covariant
\begin{eqnarray}
\Gamma_\sigma(q;P) = E_\sigma (P^2) f_0(q^2).
\end{eqnarray}

At \mbox{$T=0$} the mass-shell condition for a meson as a $\bar q q$ bound state of the BSE
is equivalent to the appearance of a pole in the $\bar q q$ scattering amplitude as a
function of $P^2$.  At $T\neq0$ in the Matsubara formalism, the $O(4)$ symmetry is broken
by the heat bath and we have \mbox{$P \to (\Omega_m,\vec{P})$} where \mbox{$\Omega_m = 2m
\pi T$}.  Bound states and the poles they generate in propagators may be investigated
through polarization tensors, correlators or Bethe-Salpeter eigenvalues.  This pole
structure is characterized by information at discrete points $\Omega_m$ on the imaginary
energy axis and at a continuum of 3-momenta. One may search for poles as a function of
$\vec{P}^2$ thus identifying the so - called spatial or screening masses for each Matsubara
mode.  These serve as one particular characterization of the propagator and the \mbox{$T >
0 $} bound states.

In the present context, the eigenvalues of the BSE become $\lambda(P^2) \to
\tilde{\lambda}(\Omega_m^2,\vec{P}^2;T)$. The temporal meson masses identified by zeros of
$1-\tilde{\lambda}(\Omega^2,0;T)$ will differ in general from the spatial masses identified
by zeros of $1-\tilde{\lambda}(0,\vec{P}^2;T)$.  They are however identical at \mbox{$T =0$}
and an approximate degeneracy can be expected to extend over the finite $T$ domain, where the
$O(4)$ symmetry is not strongly broken.

The general form of the finite $T$ pion BS amplitude allowed by the separable model is
\begin{eqnarray}
\Gamma_\pi(q_n;P_m) =\gamma_5 \left(i E_\pi (P_m^2)
  +  \gamma_4 \, \Omega_m \tilde{F}_\pi(P_m^2)
  +  \vec{\gamma} \cdot \vec{P} F_\pi(P_m^2)\right) \; f_0(q_n^2) \; .
\label{pibsaT}
\end{eqnarray}
The separable BSE becomes a $3\times 3$ matrix eigenvalue problem with a kernel that is a
generalization of Eq.~(\ref{pikernel}).  In the limit \mbox{$\Omega_m \to 0$}, as is
required for the spatial mode of interest here, the amplitude \mbox{$\hat{F}_\pi = \Omega_m
\tilde{F}_\pi$} is trivially zero.

The BS amplitude for the $\sigma$ at finite temperature has only one covariant.

The impulse approximation for the $\sigma \pi\pi$ vertex, after the extension to \mbox{$T>0$}
for spatial modes characterized by \mbox{$Q=(0,\vec{Q})$} for the $\sigma$ and
$P=(0,\vec{P})$  for the relative $\pi\pi$ momentum, takes the following form:
\begin{eqnarray}
\label{spp}
 g_{\sigma \pi\pi}(T)
&=& -2 N_c\, T \sum_n {\rm tr}_s \int \frac{d^3q}{(2\pi)^3} \,
 \Gamma_\pi(k_{n+};-\vec{P}_+)\, S(q_{n+-})\,
  \Gamma_\sigma(q_{n+};\vec{Q})\,\nonumber \\
&& \times  S(q_{n++}) \; \Gamma_\pi(k_{n-};\vec{P}_-)\; S(q_{n-})\,.
\end{eqnarray}
The corresponding width of the $\sigma$ meson is equal to
\begin{eqnarray}
\Gamma_{\sigma\rightarrow\pi\pi}(T) = \frac{3}{2}
\frac{g^2_{\sigma\pi\pi}(T)}{16 \pi M_\sigma}
\sqrt{1-\frac{4M^2_\pi(T)}{M_\sigma^2(T)}}.
\label{sigmawidth}
\end{eqnarray}

\section{Numerical analysis and conclusions}

The results for $\pi$, $\sigma$ and $\rho$ mesons as well as related
quantities are calculated with the parameters $m_0 = 6.3$ MeV, $b =
730 $ MeV, $a = 0.52$, $\Lambda_0 = 698$ MeV, $\Lambda_1 = 1.78$ GeV,
$D_0\Lambda_0^2=223$, $D_1\Lambda_1^4=137$. This set allows one to
reproduce the experimental data for $M_\pi = 140$ MeV, $f_\pi = 93$
MeV and $M_\rho = 770$ MeV and obtain the reasonable result for the
$\rho \rightarrow \pi \pi$ decay width $\Gamma_{\rho \to \pi\pi} =
228$ MeV ($\Gamma^{exp}_{\rho \to \pi\pi} = 151$ MeV). For the quark
condensate we have $\langle \bar q q \rangle^0 =
(0.212\,{\rm GeV})^3$ and the quark mass function %(\ref{quarkmassfunction})
at zero momentum
equals $m(0) = 484$ MeV.

We use this set of parameters to calculate physical observables of the scalar meson. As a
result, we obtain $M_\sigma = 762$ MeV and $\Gamma_{\sigma \to \pi\pi} =  456 $ MeV.

In this work, we consider the $T$-dependence of the quark mass function $m(p)$, $f_\pi$ and
the spatial masses in the $\pi$ and $\sigma$ channels.

\begin{figure}[h]
\centerline{ \epsfig{figure=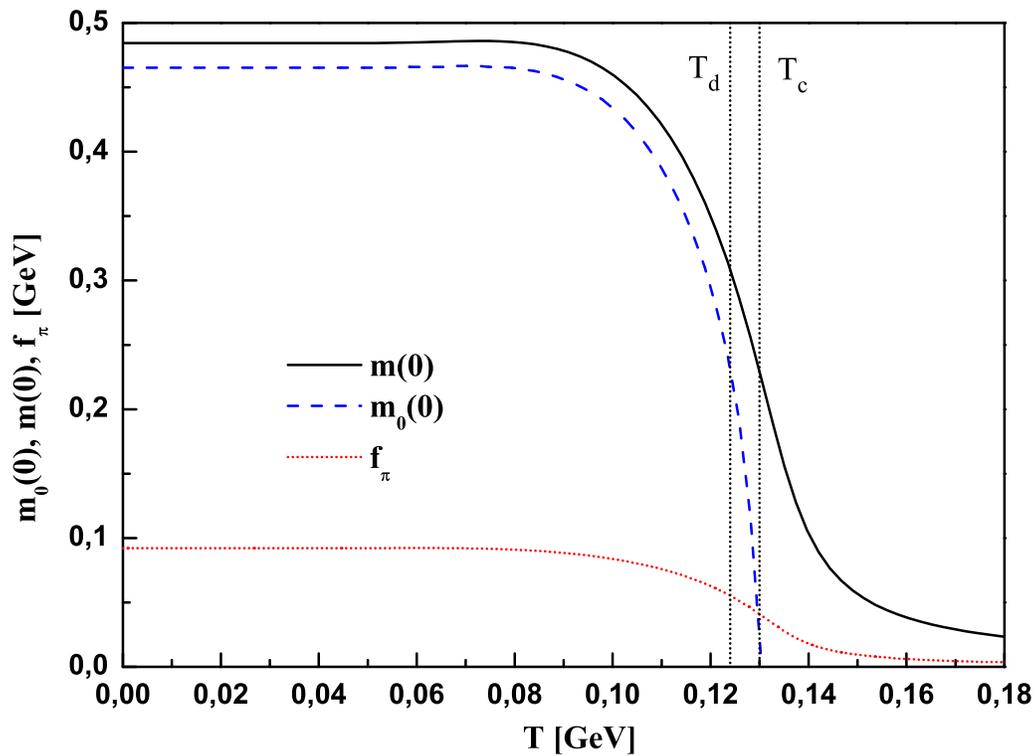,width=0.8\textwidth} } \caption{\label{fig:odin}
Temperature dependence of the quark mass function $m(0)$ (solid line). Dashed line
corresponds to the quark mass function in chiral limit, $m_0(0)$. Doted line shows  the
temperature dependence of the decay constant $f_\pi$.}
\end{figure}

The $T$-dependence of $m(0)$ , $m_0(0)$ and $f_\pi$ is displayed in Fig.~\ref{fig:odin};
$m_0(0)$ corresponds to the solution of DSE in the chiral limit $m_0=0$. The temperature DSE
equation contains three functions $A(p_n^2,T)$, $C(p_n^2,T)$ and $B(p_n^2,T)$. The numerical
calculations show that the solutions $A(p_n^2,T)$ and $C(p_n^2,T)$ as functions of
temperature practically coincide in the region $0 < T < 200$ MeV. The temperature $T_d=123$
MeV(quark deconfinement temperature) presented in Fig.~\ref{fig:odin} is determined by the
properties of Green quark function in the DSE equation. It correspods to the appearance of
the poles at real momentum in the lowest Matsubara quark mode.

The critical temperature $T_c$ is that when the quark condensate disappears and chiral
symmetry is restored.

\begin{figure}[h]
\centerline{ \epsfig{figure=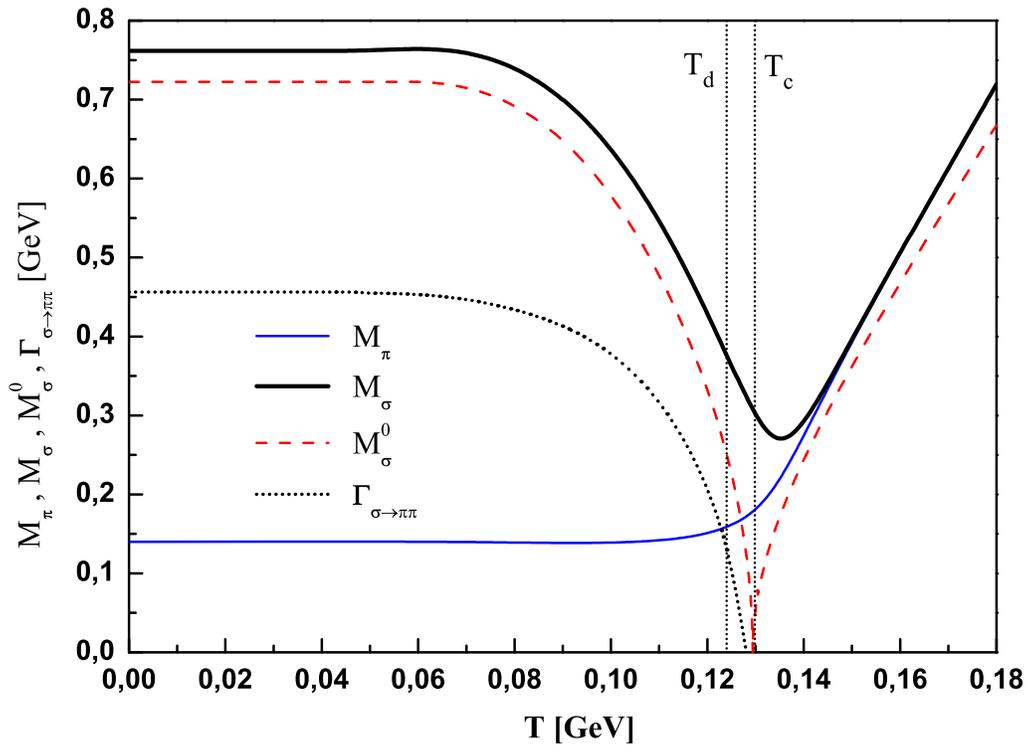,width=0.8\textwidth} }
\caption{\label{fig:dva} Temperature dependence of $M_\pi$, $M_\sigma$
and $\Gamma_\sigma$.  }
\end{figure}

%The result for the $\pi$ and $\sigma$ masses is displayed in Fig.~\ref{fig:dva}.
In the chiral limit the pion is massless below $T_c$ and its mass increases above $T_c$; the
$\sigma$ meson mass drops to zero at $T_c$ and above this temperature it is degenerated with
$M_\pi$. When $m_0 \neq 0$ $M_\pi(T)$ is seen to be only weakly $T$-dependent up to near
$T_c$, where a sharp rise begins, as displayed in Fig.~\ref{fig:dva}. These qualitative
features of the response of the pion mode with $T$ agree with the results deduced from the
DSE in \cite{cr1}-\cite{cr4}.

The temperature dependence of the width $\Gamma_{\sigma \rightarrow
\pi \pi}$  is also  shown in Fig.~\ref{fig:dva}.

In conclusion, we have studied the $\pi$ and $\sigma$ meson properties and finite temperature
in the framework of the nonlocal quark model which was used in \cite{ijmpa}. The temperature
dependence of these quantities shows that the restoration of chiral symmetry, $M_\sigma \sim
M_\pi$, occurs at the critical temperature $T_c = 130$ MeV.

%A more detailed description of the considered questions will be published in \cite{big}.

\subsection*{Acknowledgements}

%The authors thank D.~Blaschke for fruitful collaboration.
The work of Yu.L.K. has been supported by the Russian Foundation for Basic Research,
under contract 03-01-00657. A.E.R. and M.K.V. acknowledge the support the Russian
Foundation for Basic Research, under contract 05-02-16699, and of the
Heisenberg-Landau program. A.E.R. received partial financial support from Virtual
Institute "Dense Hadronic Matter and QCD Phase Transition" of the Helmholtz
Association under grant No. VH-VI-041.

%A.E.R. thanks the Theory Group of GSI for for the hospitality, where part of this work
%was conducted.

\end{document}